\documentstyle[12pt,aps]{revtex}

\include{psfig}

\input epsf

\begin{document}
\title{
\hfill\parbox[t]{2in}{\rm\small\baselineskip 14pt
{~~~~~JLAB-THY-99-30}\vfill~}
\vskip 2cm
Meson-like Baryons and the Spin-Orbit Puzzle
}

\vskip 1.0cm

\author{Nathan Isgur}
\address{Jefferson Lab, 12000 Jefferson Avenue,
Newport News, Virginia  23606}
\maketitle

\vspace{6.0 cm}

\begin{abstract}

		I describe a special class of meson-like $\Lambda_Q$ excited
states and present
evidence supporting the similarity of their spin-independent
spectra to those of mesons. I
then examine
spin-dependent forces in these baryons,
showing that predicted effects of spin-orbit forces are small for them for the same
reason
they are small for
the analogous mesons: a fortuitous cancellation
between large spin-orbit forces due to one-gluon-exchange and
equally large inverted spin-orbit forces due to Thomas precession
in the confining potential. In addition to eliminating
the baryon spin-orbit puzzle in these states, this solution 
provides a new perspective on spin-orbit forces in all baryons.

\bigskip\bigskip\bigskip\bigskip

\end{abstract}
\newpage

\section {Background}
\medskip

    Confinement in mesons and baryons should be very similar since color
dynamics is
sensitive at large distances only to the net color charge of the
interacting sources.
Thus whether a quark is bound to an antiquark or to a diquark in a color
$\bar3$ (as it must be for a baryon to be in an overall color singlet)
should make no
difference at large separation (or alternatively at high excitation).
For similar reasons, the one-gluon-exchange forces in
mesons and
baryons are very closely related:  pairwise forces in baryons  have
exactly half the strength of those in mesons (for identical spins and
separations).

    Given this close connection between meson and baryon dynamics  and the
success of the quark model in meson spectroscopy, it is surprising
that there is still an unsettled qualitative problem in the quark model for
baryons:  the so-called ``baryon spin-orbit puzzle"  that baryon spin-orbit
splittings appear to be much smaller than expected from their
one-gluon-exchange
matrix elements \cite{IK}.  However, by this criterion the mesons would
also have a spin-orbit
problem. Meson spin-orbit splittings are also much smaller than expected from
their one-gluon-exchange matrix elements \cite{LSgone,LSinversion}, but
mesons have
no spin-orbit problem because the ``normal" spin-orbit matrix element is
largely
cancelled by a  strong ``inverted" spin-orbit matrix element from
Thomas precession in the confining potential.

    	The physics behind this cancellation has received support recently
from analyses of heavy quarkonia, where both analytic techniques
\cite{Brambilla} and numerical studies using lattice QCD \cite{latticeLS}
have shown that the confining forces are spin-independent {\it apart} from
the inevitable spin-orbit pseudoforce due to Thomas precession.  Moreover,
as has been known for more than ten years, the data on charmonia  require an
inverted spin-orbit matrix element from Thomas precession in the confining
potential
to cancel part of
the strength of the  OGE matrix element \cite{LSgone}.  If the charm quark were
sufficiently massive, its low-lying spectrum would be rigorously dominated
by one gluon exchange,
and one indeed observes that the $\Upsilon$ system is closer to this ideal.
Conversely, as one moves from $c \bar c$ to lighter quarks, the
$\ell=1$
wave functions move farther out into the confining potential and the
relative strength of the Thomas precession term grows.  It is thus very
natural to expect a strong cancellation in light quark systems.

    	As shown
in the original Isgur-Karl paper on the P-wave baryons \cite{IK},
a very similar cancellation appears to occur at the two-body level
in baryons.    However, unlike mesons,
baryons can also experience three-body spin-orbit forces \cite{3bodyso} ({\it e.g.,}
potentials
proportional to $(\vec S_1-\vec S_2) \cdot (\vec r_1-\vec r_2) \times \vec
p_3$
where $\vec S_i$, $\vec r_i$, and $\vec p_i$ are the spin, position, and momentum
of quark $i$).  The
matrix elements of these three body spin-orbit forces are all calculated
in Ref. \cite {IK}, but no apparent cancellation amongst them is found.
{\it I.e.,} the
spin-orbit  puzzle might more properly be called the ``baryon
three-body spin-orbit puzzle".  In view of the facts that
one could understand the smallness of spin-orbit forces in mesons
and that the data also clearly called for small
spin-orbit forces in baryons,
the Isgur-Karl model anticipated a solution to
the baryon three-body spin-orbit puzzle  and {\it
as a
first approximation} discarded all spin-orbit forces.
It was assumed that, as in mesons, a more precise and broadly applicable
description would have to treat residual spin-orbit interactions.
In the meantime, a possible solution to this problem has been suggested
\cite{CapstickI} in which relativistic effects enhance spin-spin
over spin-orbit
effects. This suggestion may prove to be correct, though it would  then be
an accident that
in mesons a nonrelativistic solution presents itself. In this paper I
identify a special class of baryons which exhibit a meson-like solution to the
spin-orbit
puzzle and lead to a new perspective on this old problem.

\section {A Tower of Meson-like $\Lambda_Q$ Excited States}

\subsection {Introduction}
	Consider a $ud Q$ baryon in which the $ud$ quark pair is compact
and $Q$ is far
from their center-of-mass, as shown in Fig. 1(b). As previously mentioned, the $ud$ pair must be
in a color $\bar3$, so the forces between it and $Q$ are the same as those between
an antiquark and $Q$.  If the internal dynamics of the $ud$ pair were
independent
of $\vec\lambda$, then each $ud$ eigenstate would act as an extended
quasi-antiquark with which $Q$ could form a tower of meson-like excited
states.  The simplest such $ud$ pair, and the one
which is the focus of this paper, has the $ud$ pair in its isospin zero and
spin zero ground state.  I label these state $\Lambda_{Q^*}$'s since in
them only
the $Q$ relative coordinate is excited over the ground state
$\Lambda_Q{1 \over 2}^+$. We will see that the only spin-dependent
forces in these states are the spin-orbit forces experienced by
$Q$, so they are a natural choice for a system in which to investigate the
baryon
spin-orbit puzzle.

\bigskip
%
%
\begin{center}
~
\epsfxsize=2.8in  \epsfbox{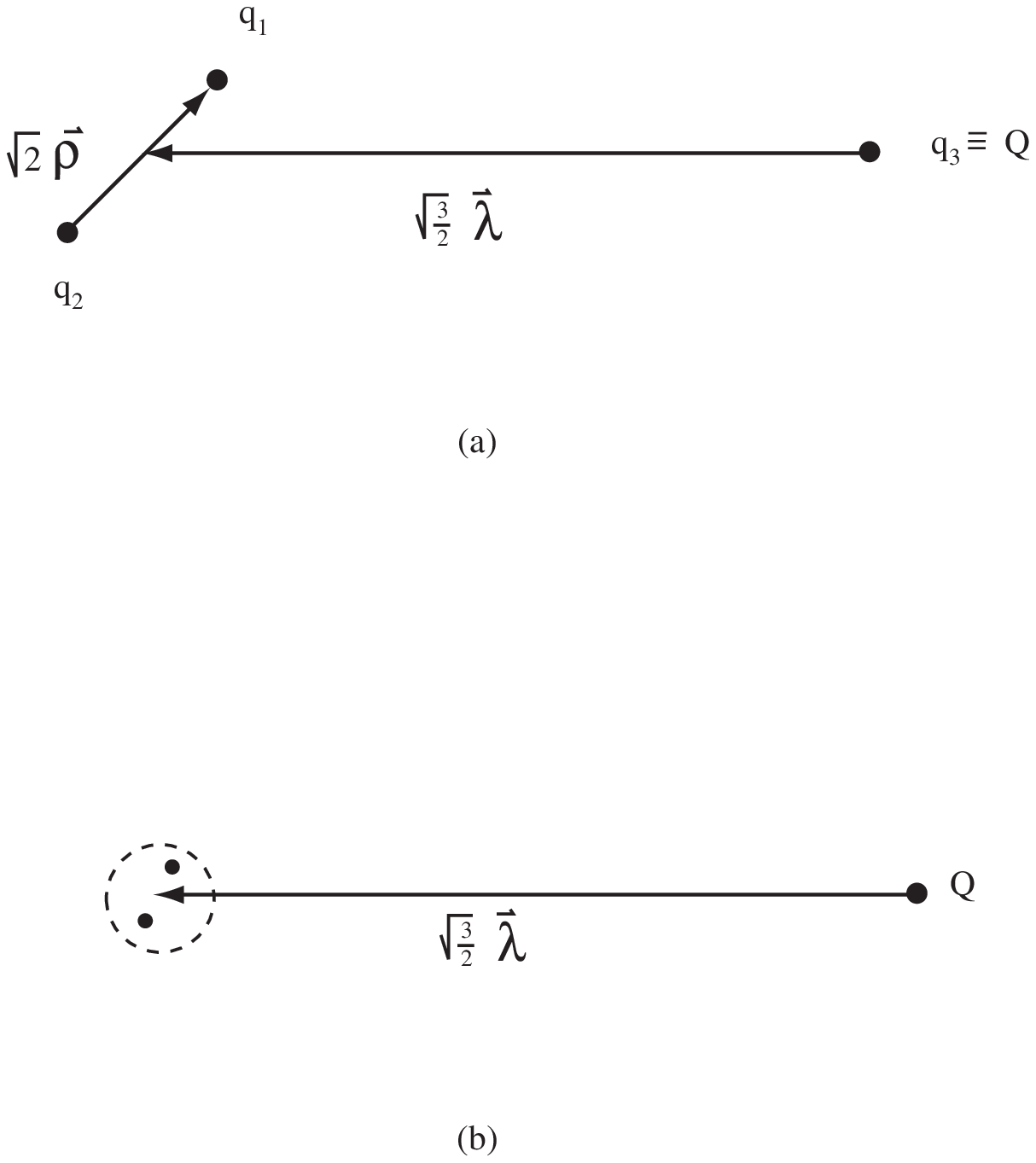}
\vspace*{0.1in}
~
\end{center}

\noindent{Fig. 1: (a) The relative coordinates 
$\vec \rho \equiv \sqrt{1 \over 2}( \vec r_1- \vec r_2)$ and
$\vec \lambda \equiv \sqrt{1 \over 6}( \vec r_1 + \vec r_2 - 2 \vec r_3)$
of a $q_1q_2q_3$ baryon with
$m_1=m_2=m$ and $m_3=m_Q$. (b) A meson-like baryon configuration.}

\bigskip\bigskip\bigskip

	The idealization of a separation between $\vec\rho$ and
$\vec\lambda$ dynamics
(see Fig. 1(a)) is actually realized in the zeroth-order states of the
Isgur-Karl (IK) model \cite{IK} where for $m_1 = m_2 = m$ and $m_3 = m_Q$
\begin{equation}
H={p_{\rho}^2 \over 2m} + {p_{\lambda}^2 \over 2m_{\lambda}} + V_{si} + V_{sd}
\end{equation}
with
\begin{equation}
m_{\lambda} \equiv {3mm_Q \over 2m+m_Q}
\end{equation}
and $V_{si}$ and $V_{sd}$ spin-independent and spin-dependent potentials.  In
the IK model one introduces an {\it artificial} harmonic term to define
\begin{equation}
V_{si}= \sum_{i<j} {1 \over 2}kr^2_{ij} + \Delta V_{si}
\label{eq:hozero}
\end{equation}
where
\begin{equation}
\Delta V_{si} \equiv  V_{si}  - \sum_{i<j} {1 \over 2}kr^2_{ij}
\label{eq:deltaV}
\end{equation}
and then treats $\Delta V_{si}$ and $V_{sd}$  as perturbations on the zeroth-order
Hamiltonian
\begin{eqnarray}
H_0&=&{p_{\rho}^2 \over 2m} + {p_{\lambda}^2 \over 2m_{\lambda}} +
\sum_{i<j} {1 \over 2}kr^2_{ij} \\
   &=&{p_{\rho}^2 \over 2m} + {p_{\lambda}^2 \over 2m_{\lambda}}
+ {3 \over 2}k \rho^2 + {3 \over 2}k \lambda^2
\label{eq:Vharmonic}
\end{eqnarray}
which separates as advertized with $\omega_{\rho}= \sqrt{3k/m}$
and $\omega_{\lambda}= \sqrt{3k/m_{\lambda}}$. 
(The choice of $\vec \rho$ and $\vec \lambda$ as variables was made
historically so that in the SU(3) limit $m_s \rightarrow m$
the $\rho$ and $\lambda$ oscillators become degenerate
with wave functions that are manifestly good representations of the permutation group $S_3$.)
Note that by construction $k$
is an auxillary parameter
which can be chosen to minimize the importance of $\Delta V_{si}$.  
In what follows I take
\begin{equation}
V_{si}= -\sum_{i<j} {2 \alpha_s \over 3r_{ij}} + c + V_{conf}
\label{eq:Vsibaryon}
\end{equation}
where the first term is the color Coulomb potential between quarks in a baryon
(to be contrasted with $- 4\alpha_s/3r$ in a meson), $c$ is a constant, and
$V_{conf}$ is the long-range confining potential.  The zeroth-order
eigenstates of (\ref{eq:Vharmonic}) are
states with spatial wave functions
$\psi_{n_{\rho} \ell_{\rho} m_{\rho}}(\vec\rho ~ ) \psi_{n_{\lambda}
\ell_{\lambda} m_{\lambda}} (\vec\lambda)$
and with the various flavor and spin states allowed by the generalized Pauli
principle (given that the quarks are in the totally antisymmetric color state
$C_A$).

	As stated earlier, I focus here on the isospin zero,
light-quark-spin-zero states $\Lambda_{Q^*}$ (i.e., those with flavor wave
function
\begin{equation}
\phi_{\Lambda}= \sqrt{1 \over 2}(ud-du)Q
\end{equation}
and spin wave functions
\begin{eqnarray}
\chi_+^{\rho}&=&\sqrt{1 \over 2}(\uparrow \downarrow - \downarrow \uparrow)
\uparrow \\
\chi_-^{\rho}&=&\sqrt{1 \over 2}(\uparrow \downarrow - \downarrow \uparrow)
\downarrow
\end{eqnarray}
as defined in Ref. \cite{IK}) with the $\vec\rho$ variable in its ground state:
\begin{equation}
\psi_{Q^*_{n_\lambda \ell_\lambda m_\lambda}}= C_A \phi_{\Lambda} \chi^{\rho}
\psi_{000}(\vec \rho ~ ) \psi_{n_\lambda \ell_\lambda m_\lambda} (\vec \lambda)
\label{eq:baryonwf}
\end{equation}
where $\psi_{n_v \ell_v m_v}(\vec v)$ is the normalized harmonic oscillator
wave function for the variable $\vec v$ with principal quantum number
$n_v$  and
angular momentum quantum numbers ($\ell_v,m_v$).  Since $\vec\lambda$ is
symmetric
under $1 \leftrightarrow 2$ interchange, these states all have the $1
\leftrightarrow 2$
antisymmetry required by the generalized Pauli principle, and the tower of
$\vec\lambda$
excited states $\Lambda_{Q^*}$ stand in one-to-one correspondence with the
states
of a $\bar\sigma Q$ meson with potential $\frac{1}{2} (2k)r^2$ where
$\vec r \equiv \vec r_{\bar \sigma} - \vec r_Q$ and $\bar \sigma$ is a ficticious
antiquark with spin and isospin zero and mass $2m$.  In the harmonic limit
each of these towers
of excited states have a spacing of $\sqrt{3k/m_{\lambda}} =
\sqrt{2k/\mu_{\sigma Q}}$
where $\mu_{\sigma Q}$ is the $\bar\sigma Q$ reduced mass.

\subsection {Beyond the Harmonic Approximation}

    As mentioned earlier, $k$ is an auxillary parameter which may be chosen
to minimize
the perturbation $\Delta V_{si}$.  Since, consistent with the $1
\leftrightarrow 2$ symmetry
of this system, Eq. (\ref{eq:hozero}) is trivially generalized to  allow
the $r_{12}$
spring constant to be distinct from the $r_{13}$ and $r_{23}$ spring constants,
the $\rho$ and $\lambda$ spring constants may be taken to be independent
auxillary parameters $k_\rho$ and $k_\lambda$.  Somewhat less trivial is
the fact that
$k_\lambda$ may be chosen independently for each value of $\ell_\lambda$ in
the tower
of $\lambda$ excitations.  The subtowers consisting of states of fixed
$\ell_\lambda$ (but excitation labelled by $n_\lambda$) are all mutually orthogonal, so
choosing $k_\lambda$
to optimize the energy and wave function of $\psi_{0  \ell_\lambda
m_\lambda} (\vec\lambda)$
is a good strategy for producing accurate orthonormalized eigenfunctions of
the
spin-independent Hamiltonian.  (I optimize $n_\lambda = 0$ for each
$\ell_\lambda$
since these are the phenomenologically most relevant states).  Given this
strategy,
one never actually resorts to perturbation theory in $\Delta V_{si}$ for
the $n_\lambda = 0$
states:  their energies and eigenfunctions are best determined
variationally.
The analogous strategy may obviously be employed for mesons.

     	In an elementary $\bar\sigma Q$ system with
\begin{equation}
V_{si}= - {4 \alpha_s \over 3r} + c' + br
\label{eq:Vsimeson}
\end{equation}
the expectation value of the spin-independent Hamiltonian
in $\psi_{0 \ell m} (r) \sim Y_{\ell m} (\theta \phi) r^\ell e^{-{1 \over
2}\beta^2 r^2}$ is
\begin{equation}
E_{0 \ell}(\beta)= ({{2 \ell+3} \over 2}) {\beta^2 \over 2 \mu_{\sigma Q}}
-{ {2^{\ell+3} \ell ! \alpha_s \beta} \over 3  (2 \ell +1)!! \sqrt{\pi}}
+c'
+{ {2^{\ell+1} (\ell +1) ! b} \over  (2 \ell +1)!!\sqrt{\pi}  \beta}~~,
\label{eq:Emeson}
\end{equation}
where $(2 \ell +1)!! \equiv (2 \ell +1)(2 \ell -1)(2 \ell - 3) \cdot \cdot \cdot 3 \cdot 1$.
Table I shows how this spectrum varies with $\mu_{\sigma Q}$.  
(Throughout this paper I use the ``standard" parameters $m_u=m_d \equiv m=0.33$ GeV,
$m_s=0.55$ GeV, $m_c=1.82$ GeV, $m_b=5.20$ GeV, and $b=0.18$ GeV$^2$, with $\alpha_s=0.6$
since we are considering large-distance dominated systems.)
In a pure
Coulomb
potential, the spacings between energy levels would grow like $\mu_{\sigma Q}$, while in a pure
linear potential they would decrease like $\mu_{\sigma Q}^{- 1/3}$. The latter behaviour
can be seen at
large $\ell$, but at low $\ell$ the
admixture of Coulomb potential
relevant to light quark spectroscopy leads to even more slowly varying splittings, as observed
in nature
(see below).

\bigskip\bigskip\bigskip

\begin{table}
  \caption { The variationally determined elementary meson spectrum (in GeV)
as a function of the reduced mass
$\mu_{\sigma Q}$.}
\vspace{0.4cm}
 \begin{center}
  \begin{tabular}{|cccccc|}
&spectral splitting  & $\mu_{\sigma Q}=m/2$ & $\mu_{\sigma Q}=m$ & $\mu_{\sigma Q}=2m$ & \\
\hline
&$E_{01}-E_{00}$ & $0.59$  & $0.53$  & $0.51$  & \\
&$E_{02}-E_{01}$ & $0.45$  & $0.38$  & $0.33$  & \\
&$E_{03}-E_{02}$ & $0.39$  & $0.32$  & $0.27$  & \\
&$E_{04}-E_{03}$ & $0.35$  & $0.28$  & $0.23$  & \\
  \end{tabular}
 \end{center}
\end{table}

\bigskip\bigskip

	We
expect the analogous
tower of  $\Lambda_{Q^*}$ baryons to have the same spectrum for sufficiently
large $\ell$.  I will now demonstrate that this is the case.  The relevant
baryon
wave functions are those of Eq. (\ref{eq:baryonwf}) with
\begin{equation}
\psi_{000}(\vec \rho ~ )={\alpha_{\rho}^{3/2} \over \pi^{3/4} } e^{- {1 \over
2} \alpha_{\rho}^2 \rho^2}
\label{eq:psirho}
\end{equation}
and
\begin{equation}
\psi_{0\ell \ell}(\vec \lambda)={{\alpha_{\lambda}^{\ell+3/2}
\lambda^{\ell}_+} \over {\pi^{3/4} }\sqrt{\ell !}}
e^{- {1 \over 2} \alpha_{\lambda}^2 \lambda^2}
\label{eq:psilambda}
\end{equation}
where $\lambda_\pm \equiv \lambda_1 \pm i\lambda_2$.  The kinetic energy
term analogous
to the first term in Eq. (\ref{eq:Emeson}) is thus
\begin{equation}
 {3 \alpha_{\rho}^2 \over 4m} + ( {2 \ell+3 \over 2}) {\alpha_{\lambda}^2
\over 2 m_{\lambda}} ~ .
\end{equation}
More interesting are the Coulomb terms analogous to the second term of Eq.
(\ref{eq:Emeson})
arising from Eq. (\ref{eq:Vsibaryon}).  The $- {2\alpha_s/3r_{12}}$ term is
straightforward, but the $- {2\alpha_s/3r_{13}}$ and $-
{2\alpha_s/3r_{23}}$ terms
have an interesting wrinkle:  Gauss' Law.  In averaging $\vec \rho$ over a
spherical
shell around the $ud$ center-of-mass, only shells that fall between that
center-of-mass and $Q$ will lead to an electric field at $Q$.  This
electric field
will be that of the charge of the shell concentrated at the center of mass,
and
will have the corresponding potential energy.  A shell that falls outside
$Q$ will
produce no electric field and a constant potential corresponding to the
potential that a
distant charge would have experienced just as it crossed the shell.  Thus
in the
special class of states we are considering here, the color Coulomb
potential takes
on the effective form
\begin{equation}
V^{eff}_{Coulomb}(\vec \rho, \vec \lambda)=V^{eff}_{\rho}(\rho)+
V^{eff}_{\lambda}(\rho, \lambda)
\end{equation}
where
\begin{equation}
V^{eff}_{\rho}(\rho)= - {2 \alpha_s \over 3 (\sqrt{2} \rho)}
\end{equation}
and
\begin{equation}
V^{eff}_{\lambda}(\rho, \lambda) = - {4 \alpha_s \over 3}[{1 \over (\sqrt{3
\over 2} \lambda)}
\theta (\lambda -{\rho \over \sqrt{3}}) +
{1 \over (\sqrt{1 \over 2} \rho)} \theta ({\rho \over \sqrt{3}}-\lambda )]~~.
\end{equation}
We may associate $V_\rho ^{eff}$ with a color electric field internal to
the $ud$ pair and
directed along $\vec\rho$ and $V_\lambda ^{eff}$ with the color electric
field described above and
directed along $\vec  \lambda$.

	After averaging over $\rho$ in the wave function
$\psi_{000} (\vec \rho ~ )$ one obtains a $\lambda$-dependent
effective potential
\begin{eqnarray}
V^{eff}_{Coulomb}(\lambda)&=& - {4 \alpha_s \alpha_\rho \over 3 \sqrt{2\pi} }
-{4 \sqrt{2} \alpha_s Q_{\alpha_\rho}(\sqrt{3\over 2} \lambda) \over 3 \sqrt{3} \lambda }
- {16 \alpha_s \alpha_\rho \over 3 \sqrt{2\pi} }e^{-3\alpha^2_\rho \lambda^2} \\
&& \nonumber \\
&=& - {4 \alpha_s \alpha_\rho \over 3 \sqrt{2\pi} }
-{4 \sqrt{2} \alpha_s ~ erf(\sqrt{3} \alpha_\rho \lambda) \over 3 \sqrt{3} \lambda }
\end{eqnarray}
where
\begin{equation}
 Q_{\alpha_\rho}(\sqrt{3\over 2} \lambda) \equiv {\alpha^3_{\rho} \over \pi^{3/2}}
\int_0^{\sqrt{3} \lambda} d^3\rho~e^{-\alpha^2_\rho \rho^2}
\end{equation}
is the ``charge" inside the radius $\sqrt{\frac{3}{2}}$ $\lambda$ and
$erf(z) \equiv {1 \over \sqrt{\pi}}\int_{-z}^z dx e^{-x^2}$.  Notice that
\begin{equation}
E^{eff}_{Coulomb}(\lambda) \equiv - {dV^{eff}_{Coulomb} \over d(\sqrt{3
\over 2}\lambda)}
= -{4  \alpha_s Q_{\alpha_\rho}(\sqrt{3\over 2} \lambda) \over 3 (\sqrt{3 \over 2}
\lambda)^2 }
\end{equation}
as required, and that the energy associated with the Coulomb potentials is the
expectation value of $V_{Coulomb}^{eff} (\lambda)$ in the wave function
$\psi_{0\ell\ell} (\vec \lambda)$.  Due to the appearance of 
$Q_{\alpha_\rho}(\sqrt{3\over 2} \lambda)$ 
this energy cannot be displayed in closed form, and
variational solutions must be found numerically.  However, since
 $\langle \lambda^2 \rangle^{1/2} = \sqrt{\ell+{3 \over
2}}~/\alpha_\lambda$
and since the $\alpha_{\lambda}$ which
minimizes the energy decreases as $\ell$ increases, the Coulomb energy quickly
``heals" to the meson-like value
\begin{equation}
-{ {4 \alpha_s  \alpha_\rho} \over 3   \sqrt{2\pi}}
-{ {2^{\ell+3} \ell ! \alpha_s (\sqrt{2\over 3} \alpha_\lambda)} \over 3
(2 \ell +1)!! \sqrt{\pi}}
\end{equation}
where the first term is the $ud$ energy and $\sqrt{\frac{2}{3}}
\alpha_\lambda$ is the
appropriate analog of $\beta$ since $\sqrt{\frac{3}{2}} \lambda$
corresponds to $r$. In practice this ``healing" occurs very
rapidly: in  the $\Lambda_{s^*}$ baryons we will
consider below, the Coulomb energy differs from its meson-like value by only
$5\%$ in $\ell=1$ and $2\%$ in $\ell=2$.

	     Finally, we consider the confining potential.  From the
preceding discussion we
know that the color Coulomb field in the special class of states we are
considering here
has two components:  one of magnitude $-{2\alpha_s /
3(\sqrt{2}\rho)^2}$
internal to the $ud$ pair which is directed along $\vec \rho$, and one of
asymptotic magnitude $-{4 \alpha_s /  3 (\sqrt{\frac{3}{2}}\lambda)^2}$
between the $ud$ pair
and $Q$ which is directed along $\vec\lambda$.  Since confinement evolves
out of the color electric field at large distances, it is  natural to
assume, in analogy
to the standard meson hypothesis encapsulated in Eq. (\ref{eq:Vsimeson}),
that in these
states
\begin{equation}
V^{eff}_{conf}(\rho,\lambda)={1 \over
2}b(\sqrt{2}\rho)+b(\sqrt{\frac{3}{2}}\lambda)
\end{equation}
up to an overall constant.
Since the second term is the analog of the meson confining potential $br$,
the demonstration
that at sufficiently large $\ell$ our tower of $\Lambda_{Q^*}$ baryons will
have the
same  spectrum as the $\bar\sigma Q$ mesons is complete.

	In summary, I have shown in this Section that in the variational
wave functions
(\ref{eq:baryonwf}), and {\it a fortiori} in lowest order perturbation
theory in $\Delta V_{si}$ of Eq. (\ref{eq:deltaV}), a tower of
$\Lambda_{Q^*}$ baryons
with the $ud$ pair in $\ell_\rho=0$ and spin zero exists which will be
analogous to the
spectrum of a ficticious meson $\bar \sigma Q$ containing a scalar
antiquark of mass $2m$.
Gauss' Law has produced a situation that is only slightly more complicated
than the
idealized harmonic limit wherein the $\vec\rho$ and $\vec\lambda$ variables
completely separate:  they have become coupled only through the effect of
$\psi_{000}(\rho)$ on the effective charge $Q_{\alpha_\rho}(\sqrt{3\over 2} \lambda)$.
Physically this means that the size of the $ud$ pair's wave function is not
fixed, but rather that it grows to an asymptotic value as $\ell_\lambda$
increases.

\subsection {An Empirical Meson-Baryon Correspondence}

    Though there are no mesons with scalar antiquarks $\bar\sigma$, the
spin-independent quark model 
spectra of such mesons would be the same as those of ordinary $\bar q Q$ mesons
with $m_q=m_\sigma$.  Moreover, Figs. 2 and 3 show what Table I has led us to
anticipate:  the spectra of mesons are slowly-varying functions of their
reduced
masses.  Thus we may reasonably look for a correspondence
between the spectrum of any flavor of
mesons with
reduced mass of the order of $m$ with the select tower of $\Lambda_{Q^*}$
baryons
associated with any heavy quark $Q$.  The most extensive data exists for
the $K^*$ and the ordinary
$\Lambda^*$ (i.e., $\Lambda^*_s$) states, as it happens, and this
correspondence is shown for the ``fully stretched" total angular momentum states
of these systems in Fig. 4.  
I first note that a comparison of the observed splittings in the $K^*$ system
with those of Table I indicates that the framework adopted here is in fact quite reasonable.
Since the $K^*-\Lambda_{s^*}$ correspondence should be best for high $\ell$,
I have aligned the highest $\ell$ established for both spectra, namely $\ell=3$
corresponding to a $4^+$ $K^*$ and a ${7 \over 2}^-$ $\Lambda_{s^*}$. The excellent correspondence
at high $\ell$ provides a detailed
view in the context of the quark model of the well-known relation between
the slopes of meson and baryon Regge trajectories. That the
$\ell =0 \rightarrow \ell = 1$ and $\ell = 1 \rightarrow \ell = 2$
splittings are smaller in the $\Lambda_{s^*}$'s than in the $K^*$'s is what
we expect from the analysis of the previous Section:  for low $\ell$ the
Coulomb potential is weakened since the $ud$ color charge is distributed
in the wave function $\psi_{000}(\vec\rho ~ )$.  Figure 4 therefore also shows
the spectrum of a ficticious $\tilde \Lambda_{s^*}$ system obtained from the experimental
$\Lambda_{s^*}$ spectrum by adding to it the perturbation that would occur if
the spatial extension of the  $ud$ pair were set to zero, demonstrating
that even the low $\ell$ behavior of the $K^*$ and $\Lambda_{s^*}$ systems are
related as expected.

\bigskip\bigskip
%
%
\begin{center}
~
\epsfxsize=2.6in  \epsfbox{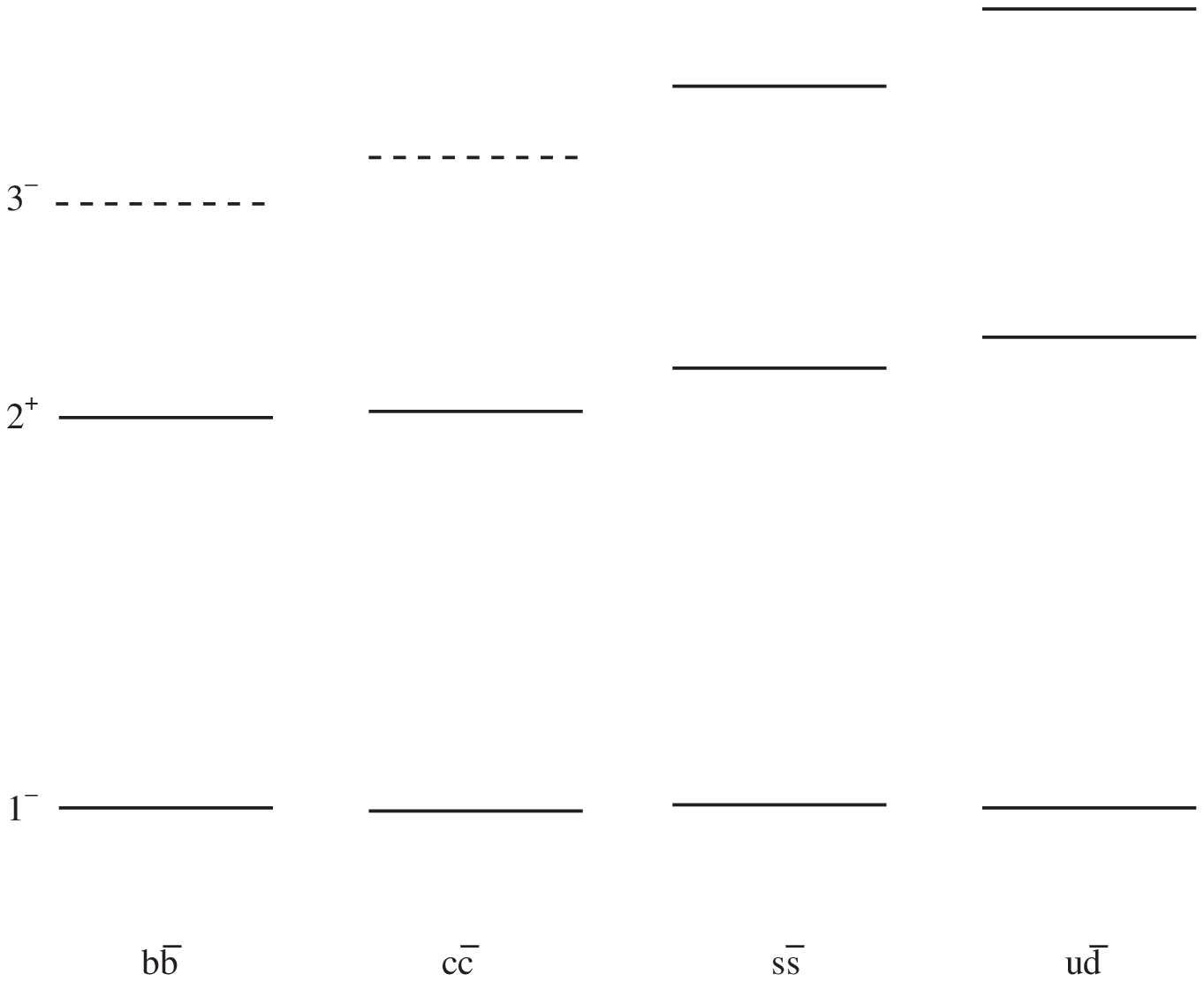}
\vspace*{0.1in}
~
\end{center}

\noindent{Fig. 2: The orbital excitation spectra of the quarkonia $Q \bar
Q$ as a
function of $m_Q$; the $1^-$ $S$-waves have
been aligned to display the splittings to the
$P$-wave $2^+$ and $D$-wave $3^-$ states. In $b \bar b$ and $c \bar c$,
the $D$-wave positions
shown as dashed lines are predictions of reliable heavy quarkonium calculations. 
The spectra are shown to scale, which may conveniently be
calibrated from the $\chi_{c2}-\psi$ splitting
of 459 MeV.}

%
%
\begin{center}
~
\epsfxsize=2.5in  \epsfbox{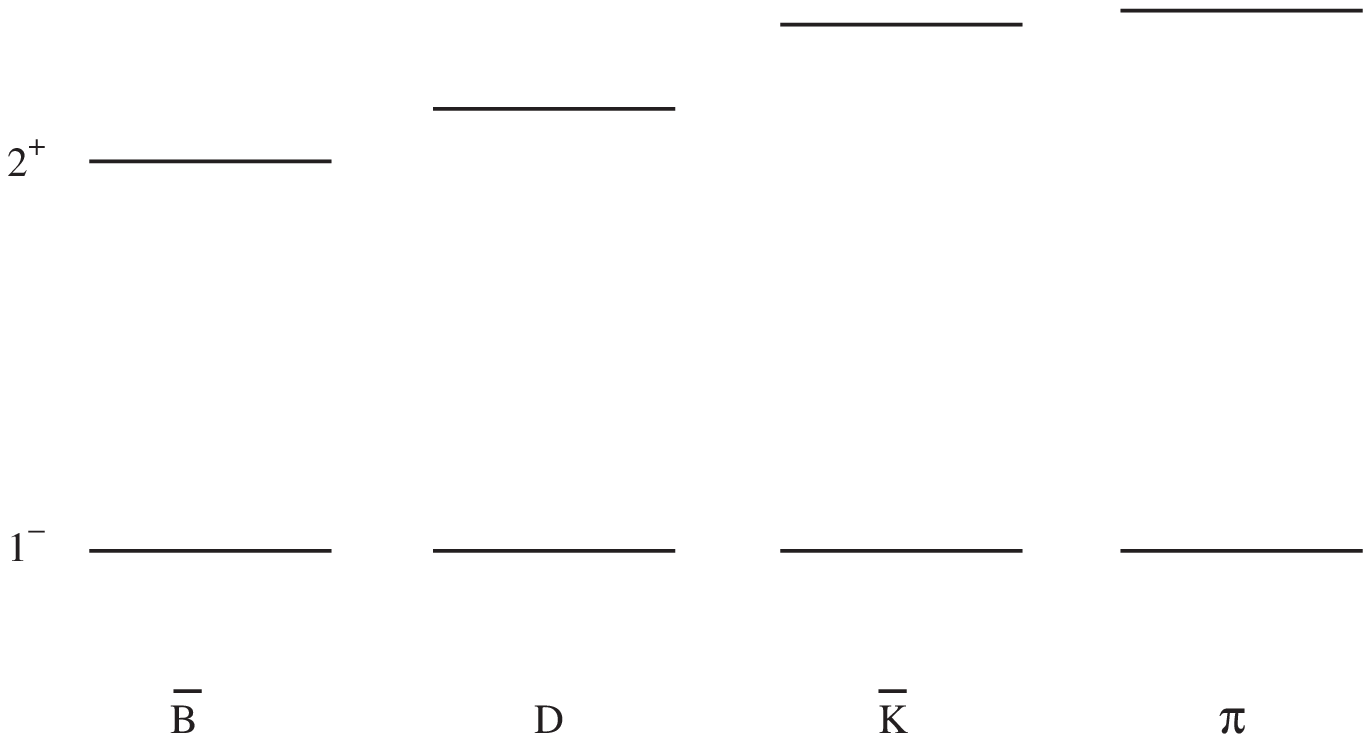}
\vspace*{0.1in}
~
\end{center}

\noindent{Fig. 3: The orbital excitation spectra of heavy-light mesons  as a
function of $m_Q$; the $1^-$ $S$-waves have
been aligned to display the splittings to the
$P$-wave $2^+$ states. The spectra are shown to scale, which may conveniently be
calibrated from the $D^*_2-D^*$ splitting
of 452 MeV.}

\bigskip\bigskip\bigskip

\bigskip
%
%
\begin{center}
~
\epsfxsize=2.6in  \epsfbox{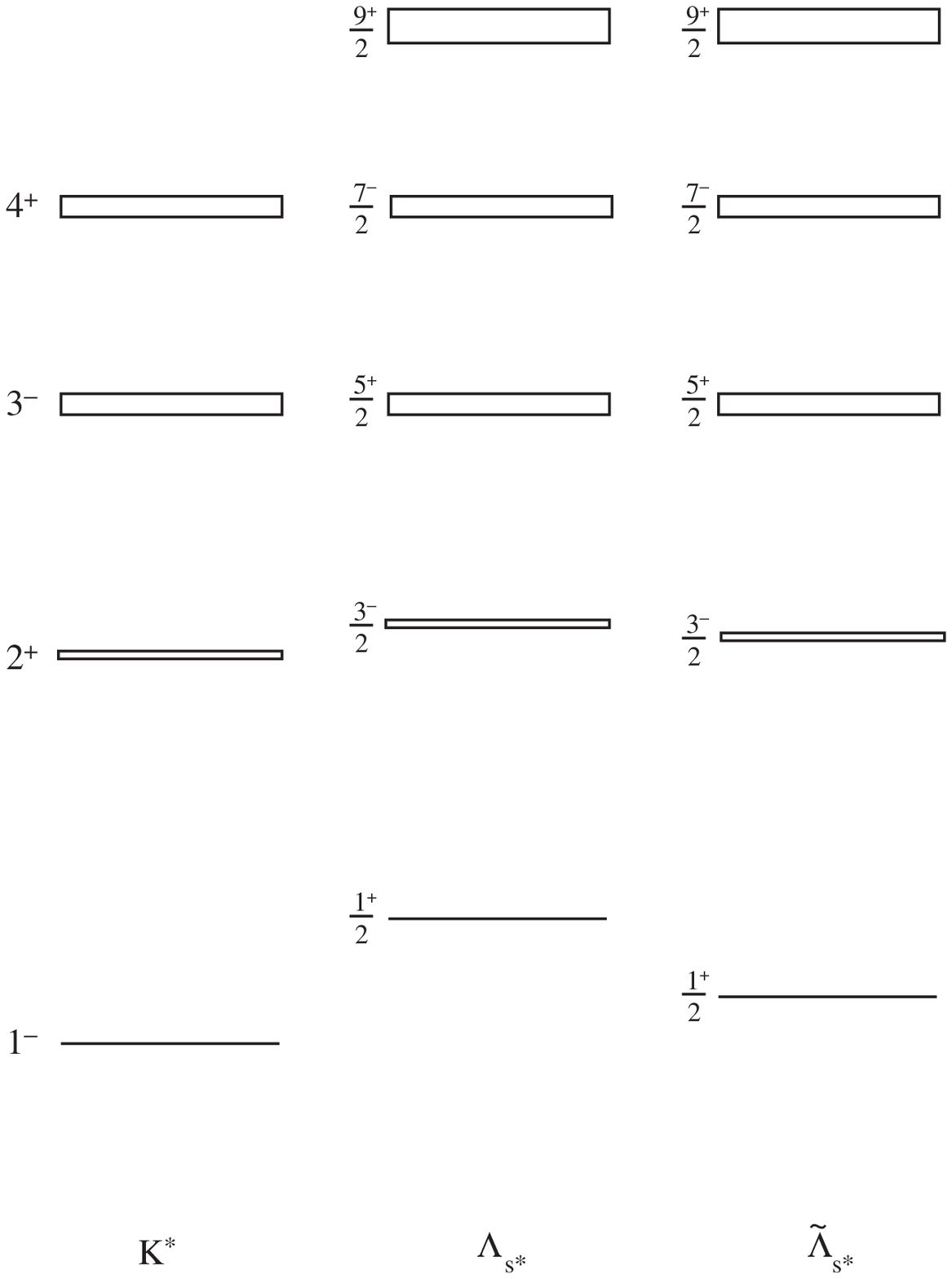}
\vspace*{0.1in}
~
\end{center}

\noindent{Fig. 4: Comparison of the spectra of well-established $K^*$ mesons and
$\Lambda_{s^*}$ baryons. The boxes shown represent the experimental
uncertainties in the masses of states. The
$\ell=3$ states have been aligned as described in the text. The third spectrum is that of a
fictitious $\tilde \Lambda_{s^*}$
baryon with the spatial distribution of the $ud$ pair set to zero. 
The spectra are shown to scale, which may conveniently be
calibrated from the $K^*_2-K^*$ splitting
of 534 MeV.}

\bigskip\bigskip\bigskip

\section {Spin-Orbit Forces in $\Lambda_{Q^*}$ Baryons}

    Given the remarkable correspondence between the spin-independent spectra
of mesons and the $\Lambda_{Q^*}$ baryons, this special tower of states seems
ideal for investigating the baryon spin-orbit puzzle.  Since the color
dynamics of these two systems are so similar, and since there is no meson
spin-orbit puzzle, one would naively expect to find no problem with spin-orbit
forces in the $\Lambda_{Q^*}$ states.  This is in fact what I find.

\subsection {A Review of Meson Spin-Orbit Forces}
	I begin with a review of meson spin-orbit forces.  In a ``real"
$\bar q Q$ meson with two
spin-$1/2$ particles, the spin-dependent potential is
\begin{equation}
V^{\bar q Q}_{sd} = V^{\bar q Q}_{ss} + V^{\bar q Q}_{so}
\end{equation}
where $V_{ss}^{\bar q  Q}$ is the usual spin-spin interaction (consisting
of the Fermi contact term and the tensor interaction) and
\begin{equation}
V^{\bar q Q}_{so} = {4 \alpha_s \over 3 \mu_{\bar q Q} r^3} \vec L \cdot
({\vec S_Q \over m_Q} +
{\vec S_{\bar q} \over m_q}) - ({2 \alpha_s \over 3 r^3}+{b \over 2r})
\vec L \cdot ({\vec S_Q \over m^2_Q} +
{\vec S_{\bar q} \over m^2_q})~~.
\label{eq:Vsomeson1}
\end{equation}
Here the first term is the dynamic spin-orbit interaction arising from
the interaction of the color magnetic moments ${\vec S_Q / m_Q}$ and ${\vec
S_{\bar q}/ m_q}$
with the color magnetic fields generated by the motion of the other quark and the second
term is the kinematic spin-orbit effect arising from Thomas precession in
the central spin-independent potential.  One can equivalently write
\begin{equation}
V^{\bar q Q}_{so} = {2 \alpha_s \over 3  r^3} \vec L \cdot [\vec S_Q ({1
\over m^2_Q} + {2 \over m_Q m_q})
+\vec S_{\bar q} ({1 \over m^2_q} + {2 \over m_Q m_q})] - {b \over 2r}
\vec L \cdot [{\vec S_Q \over m^2_Q} +
{\vec S_{\bar q} \over m^2_q}]
\label{eq:Vsomeson2}
\end{equation}
in which form the matrix elements of $1/r^3$ and $1/r$ are separated,
corresponding to the complete one-gluon-exchange component and the complete
confinement component of the spin-orbit interaction, respectively.  In the
$m_Q = m_q = m$ isovector meson sector
\begin{equation}
V^{\bar q Q}_{so} \rightarrow V^{\bar d u}_{so} = \vec L \cdot \vec S ~ [{2
\alpha_s \over m^2  r^3}
 - {b \over 2m^2 r}]
\label{eq:Vsoisovector}
\end{equation}
where $\vec S = \vec S_Q + \vec S_{\bar q} = \vec S_u + \vec S_{\bar d}$.  When
evaluated in the $\ell = 1$ wave functions of the spin-independent potential
(\ref{eq:Vsimeson}), the one-gluon-exchange component has a matrix element
of +180 MeV, while
the Thomas precession component has a matrix element of -170 MeV.
(These matrix elements are multiplied by $\langle \vec L \cdot \vec S \rangle$
which is $+1$, $-1$, $-2$, and $0$ in the $a_2$, $a_1$, $a_0$, and $b_1$,
respectively.)  The total matrix element of Eq. (\ref{eq:Vsoisovector}) can be extracted from the mixture of
contact, tensor, and spin-orbit terms contributing to the experimental P-wave masses from
the combination $\frac {5} {12} a_2 - \frac {1}{4} a_1 - \frac {1}{6} a_0$ and is 
$-3 \pm {20} $ MeV.
We see that the mesons would have a serious spin-orbit problem if Thomas
precession
in the confining potential were ignored, but that the sum of the
one-gluon-exchange
spin-orbit forces and confining Thomas precession spin-orbit forces
leads to a very small net
spin-orbit splitting as observed.

	In heavier quarkonia $Q \bar Q$, as $m_Q$ increases $\langle
1/m_Q^2 r^3 \rangle$
decreases (in a linear potential it would decrease exactly like
$m_Q^{-1}$), but the
ratio of the matrix element of $1/r$ to that of $1/r^3$ also decreases
(like $m^{-2/3}$ in a linear potential) so that very heavy quarkonia are
totally
dominated by one-gluon-exchange spin-orbit forces.  As previously mentioned, this decreasing
importance of
Thomas precession in the confining potential is observed in the $\chi_c$
and $\chi_b$ states.

	In heavy-light mesons $\bar d Q$ in the heavy quark limit $m_Q
\rightarrow \infty$,
\begin{equation}
V^{\bar q Q}_{so} \rightarrow V^{\bar d Q}_{so}\vert_{m_Q \rightarrow
\infty} =
\vec L \cdot \vec S_{\bar d} ~ [{2 \alpha_s \over 3 m^2  r^3}
 - {b \over 2m^2r}]~~.
\label{eq:Vsohqlimit}
\end{equation}
It is this interaction which determines the splitting between the heavy
quark spin
multiplets \cite{IW} with $s_{\ell} ^{\pi_\ell} = 3/2^+$ and $s_{\ell}^{\pi_\ell} =
1/2^+$ associated with
the $\ell =1$ excitations of the $\bar d Q$ system in the quark model.
Since the wave
function parameter $\beta$ (see Eq. (\ref{eq:Emeson})) increases by only
about $25\%$  
between $\bar d u$ and
$\bar d Q$,  it is to be expected
\cite{LSinversion} from
comparing Eqs. (\ref{eq:Vsohqlimit}) and (\ref{eq:Vsoisovector}) that these
two multiplets will
be inverted.  Unfortunately, no $s_{\ell}^{\pi_\ell} = 1/2^{+}$ charm or
beauty mesons
are known, so this expectation is untested experimentally \cite{Lewis}.  The spin-orbit
splitting {\it inside}
the $s_{\ell}^{\pi_\ell} = 3/2^+$ multiplet is controlled in leading order
in $1/m_Q$ by
the pure one-gluon-exchange operator
\begin{equation}
{4 \alpha_s \over 3 m m_Q  r^3}\vec L \cdot \vec S_Q~~.
\label{eq:mesononeovermQ}
\end{equation}
This operator produces a predicted splitting between $D_{2}^*$ and $D_1$ of the charm
$s_{\ell}^{\pi_\ell} = 3/2^+$
spin multiplet of $50$ MeV, comparable to the observed splitting of $45$
MeV.  Spin-spin
interactions can also contribute to this splitting at this order in
$1/m_{Q}$, but their
calculated effect is very small ($\simeq 5$ MeV).

	In summary, mesons not only have no spin-orbit problem, but their
splittings are
reasonably well described by the standard spin-orbit interaction
(\ref{eq:Vsomeson1}).

\subsection {The $\Lambda_{Q^*}$ Baryons: Theory}

    The meson-like tower of $\Lambda_{Q^*}$ baryons is actually simpler
than the analogous
$\bar dQ$ mesons because with only the spin $\vec S _Q$ of the heavy quark
active,
spin-orbit forces are the only spin-dependent forces in first order
perturbation theory. This is because  in the $\Lambda_{Q^*}$ states the  $ud$ spin-zero wave
function factorizes
from the dynamical parts of the wave function and $\langle \vec S_1 \rangle$ and
$\langle \vec S_2 \rangle$
are zero, leaving as the only operative spin-dependent interaction
\begin{equation}
V^{\Lambda_{Q^*}}_{so} = \Bigl[ {1 \over m_Q \mu_{\sigma Q} (\sqrt{3 \over
2}\lambda)}
{dV^{eff}_{Coulomb} \over d (\sqrt{3 \over 2}\lambda)}
-{1 \over m^2_Q} \Bigl(
{1 \over 2(\sqrt{3 \over 2}\lambda)}{dV^{eff}_{Coulomb} \over d (\sqrt{3
\over 2}\lambda)}
+{b \over 2(\sqrt{3 \over 2}\lambda)} \Bigr) \Bigr]
\vec L_{\lambda} \cdot \vec S_Q~~,
\label{eq:VsoLambdaQ}
\end{equation}
where the first and second terms, in an obvious notation, are 
$V^{\Lambda_{Q^*}}_{so(dynamic)}$ and $V^{\Lambda_{Q^*}}_{so(Thomas)}$.
In the
heavy quark limit \cite{IW} $m_Q \rightarrow \infty$ this interaction
becomes in leading
order in $1/m_Q$ simply
\begin{equation}
{1 \over m_Q \mu_{\sigma Q} (\sqrt{3 \over 2}\lambda)}
{dV^{eff}_{Coulomb} \over d (\sqrt{3 \over 2}\lambda)}
\vec L_{\lambda} \cdot \vec S_Q ~~,
\label{eq:VsoLambdaheavyQ}
\end{equation}
the operator analogous to that responsible for the $D_2^* - D_1$ splitting.
In the
very crude approximation of taking the $\ell = 1$ $D$ and $\Lambda_{c^*}$ wave
functions to be identical, approximating ${dV_{Coulomb} ^{eff} / d(\sqrt{3
\over 2} \lambda)}$ by
$4{\alpha_s}/3(\sqrt{3\over 2}\lambda)^2$, working to leading order in
$1/m_c$, and ignoring
spin-spin interactions in the $D_2 ^* - D_1$ splitting, the 
$D^*-\Lambda _{Q^*}$ analogy would
lead to the result
\begin{equation}
m_{\Lambda_c {3 \over 2}^-}-m_{\Lambda_c {1 \over 2}^-}\simeq {1 \over
2}(m_{D_2^*}-m_{D_1})
\simeq 20~ {\rm MeV}
\nonumber
\end{equation}
in reasonable agreement with the observed splitting of $30$ MeV \cite{PDG}.
If we
go further and use $1/m_Q$ scaling down to  $m_s$, the observed
$\Lambda_c {3 \over 2}^- - \Lambda_c {1 \over 2}^-$  splitting would lead to
$\Lambda{3 \over 2}^- - \Lambda {1 \over 2}^- \simeq 110$ MeV, in
unreasonably good agreement with the observed
$\Lambda(1520) {3 \over 2}^- - \Lambda(1405){1 \over 2}^-$ splitting
of $115$ MeV.    I will show below that these crude estimates are not that
misleading,
but from them one can already begin to anticipate the announced conclusion that there
will also be no spin-orbit
problem in the $\Lambda_{Q^*}$ baryons.

	This conclusion appears to be inconsistent with that of
Ref.\cite{IK} where two-body
spin-orbit forces invited a meson-like
cancellation, but three-body spin-orbit forces did not.  We will see in what
follows that the inconsistency is only apparent:  meson-like behaviour of
the $\Lambda_{Q^*}$
states implies that the three-body spin-orbit forces have a structure which
leads to an overall
``quasi-two-body" spin-orbit force between $Q$ and the ``quasi-antiquark"
$\bar \sigma$ made of
$u$ and $d$.  

	In the picture I have adopted here, in which the $ud$ pair is
treated as an extended scalar
antiquark $\bar \sigma$ of mass $2 m$, the meson-like character of
spin-orbit forces seems
obvious.  However, given its importance, we now examine this conclusion very
carefully from a three-body perspective.
The essential issues are all present in the simplified case where
spin-dependent
forces perturb zeroth-order
harmonic oscillator states (\ref{eq:baryonwf}), and so it is this
case I will discuss
explicitly.  In the absence
of degeneracies
in the spectrum, $V_{sd}$ will
perturb the energy of these $\Lambda _{Q^*}$ states by
\begin{equation}
\Delta E_{n_\lambda \ell_\lambda} = \langle \psi_{Q^*_{n_\lambda
\ell_\lambda \ell_\lambda}} \vert
V_{sd} \vert \psi_{Q^*_{n_\lambda \ell_\lambda \ell_\lambda}} \rangle ~~.
\end{equation}
First consider a term of $V_{sd}$ of the form $\vec S_1 \cdot \vec V$ where
$\vec V$ is a vector
operator formed from the other variables of the three body system.  Since,
$\langle \chi_{\pm}^{\rho} \vert \vec S_1 \vert \chi_{\pm}^{\rho} \rangle =0$,
such a term cannot
contribute to $\Delta E_{n_{\lambda}\ell_{\lambda}}$.  Similarly,
$\vec S_2 \cdot \vec V$ terms  do not contribute, so
as described above only $\vec S_Q \cdot \vec V$ - type terms with $\vec V$
independent
of $\vec S_1$ and $\vec S_2$, i.e.,
spin-orbit interactions of $Q$, can contribute.   The one-gluon-exchange part of these
interactions is
\begin{equation}
V_{so}^{oge}= \sum_{i=1,2} {2 \alpha_s \over 3 r^3_{iQ}} {\vec S_Q \over
m_Q} \cdot
(\vec r_{iQ} \times [{\vec p_i \over m}-{\vec p_Q \over m_Q}]+{1 \over
2m_Q}\vec r_{iQ} \times \vec p_Q)
\end{equation}
where the first and second terms are its dynamic and Thomas precession
pieces, respectively.

	I begin with the Thomas precession piece $V^{oge}_{so(Thomas)}$.  The
$\Lambda_{Q^*}$ states
of definite $jm$  are of the
form $\vert \Lambda_{Q^*_{jm}} \rangle
=C_{\ell_{\lambda}m_{\lambda}\sigma}^{jm}
\vert \psi_{Q^*_{n_{\lambda}\ell_{\lambda}m_{\lambda}}} Q_{\sigma} \rangle$
where $\sigma = \pm$ and the $C_{\ell_{\lambda}m_{\lambda}\sigma}$ are
simple Clebsch-Gordan coefficients.
In these states the expectation value of $V^{oge}_{so(Thomas)}$ is
\begin{eqnarray}
\langle \Lambda_{Q^*_{jm}} \vert  V^{oge}_{so(Thomas)}   \vert \Lambda_{Q^*_{jm}}
\rangle &=&
C_{\ell_{\lambda} m'_{\lambda} \sigma '}^{jm~~*} C_{\ell_{\lambda}
m_{\lambda} \sigma}^{jm}
\sum_{i=1,2} \langle \sigma ' \vert  \int d^3 \lambda \int d^3 \rho
\nonumber \\
&&
~~~~~\psi^*_{n_{\lambda}\ell_{\lambda}m'_{\lambda}}(\vec \lambda)
\psi^*_{000}(\vec \rho ~ )
{\alpha_s \vec S_Q \cdot \vec r_{iQ} \times \vec p_Q \over 3m_Q^2 r^3_{iQ}}
\psi_{000}(\vec \rho ~ ) \psi_{n_{\lambda} \ell_{\lambda} m_{\lambda}}(\vec
\lambda) \vert \sigma \rangle   \\
&=&C_{\ell_{\lambda}m_{\lambda}\sigma
'}^{jm~~*}C_{\ell_{\lambda}m_{\lambda}\sigma}^{jm}
\sum_{i=1,2} \langle \sigma ' \vert  \int d^3 \lambda \nonumber \\
&&
~~~~~\psi^*_{n_{\lambda}\ell_{\lambda}m_{\lambda}}(\vec \lambda)
{ \vec S_Q \cdot \vec E_{iQ}(\vec \lambda) \times \vec p_Q \over 2m_Q^2}
\psi_{n_{\lambda}\ell_{\lambda}m_{\lambda}}(\vec \lambda) \vert \sigma
\rangle
\end{eqnarray}
since $\vec S_Q$ and $\vec p_Q = - \sqrt{2 \over 3} \vec p_\lambda$ are
independent of
$\vec \rho$ \cite{oporder}, where
\begin{equation}
\vec E_{iQ}(\vec \lambda)=\int d^3 \rho ~\vert \psi_{000}(\vec \rho ~ ) \vert^2
{2 \alpha_s \vec r_{iQ} \over 3 r^3_{iQ}}
\label{eq:EQ}
\end{equation}
is the color electric field at the position of $Q$ due to the quark $i$ in
the spherically symmetric wave
function $\psi_{000}(\vec \rho ~ )$. Thus
\begin{equation}
\langle \Lambda_{Q^*_{jm}} \vert  V^{oge}_{so(Thomas)}   \vert \Lambda_{Q^*_{jm}}
\rangle=
\langle \Lambda_{Q^*_{jm}} \vert  V^{oge~\Lambda_{Q^*}}_{so(Thomas)}   \vert
\Lambda_{Q^*_{jm}} \rangle
\end{equation}
where
\begin{equation}
V^{oge~\Lambda_{Q^*}}_{so(Thomas)} = - ~ {1 \over 2m_Q^2(\sqrt{3 \over 2} \lambda)}
{dV^{eff}_{Coulomb} \over d(\sqrt{3 \over 2} \lambda)} \vec L_{\lambda}
\cdot \vec S_Q
\end{equation}
as advertized in Eq. (\ref{eq:VsoLambdaQ}).  The dynamical piece is
somewhat more complex.  Since,
for example,
\begin{eqnarray}
\vec r_{1Q} \times [ {\vec p_1 \over m}-{\vec p_Q \over m}] &=&
(\sqrt{3 \over 2}\vec  \lambda + \sqrt{1 \over 2} \vec \rho) \times
(\sqrt{3 \over 2} {\vec p_\lambda \over m_\lambda} + \sqrt{1 \over 2} {\vec
p_\rho \over m})  \nonumber \\
&=&{3 \over 2 m_\lambda}\vec L_{\lambda}+{1 \over 2m}\vec L_{\rho}
+ {\sqrt{3} \over 2 m_\lambda} \vec \rho \times \vec p_\lambda
+ {\sqrt{3} \over 2 m} \vec \lambda \times \vec p_\rho~~~,
\label{eq:3bodyso}
\end{eqnarray}
we must consider the expectation value analogous to that shown above with
$\vec r_{i Q} \times \vec p_Q$
replaced by each of these four operators.  Consider first $\vec L_\rho$:  since
$\vec L_\rho \psi_{000}(\vec \rho ~ ) = 0$,
it vanishes immediately.  Next consider $\vec \rho \times \vec p_{\lambda}$.
Since $\vec p_{\lambda}$ can be removed from the
$\vec \rho$ integration, and since this integration can be organized in
annular
regions around the direction of  $\sqrt \frac{3}{2} \vec \lambda$ which hold
$\vert \vec r_{1Q} \vert$ fixed,
$\int d^3 \rho ~\vert \psi_{000}(\vec \rho ~ )\vert^2 {\vec \rho / r^3_{1Q}} =0$.
Similarly, since  $\vec p_\rho$ operating on
$\psi_{000} (\rho)$ gives a result proportional to $\vec \rho
\psi_{000}(\rho)$,
the  $\vec \lambda \times \vec p_{\rho}$ term vanishes.
Thus the net effect of the dynamic term is carried by the  ${3 \vec
L_\lambda / 2m_\lambda} $ term
(which is identical for $i=1$ and 2).
Using the preceding, one easily obtains
\begin{equation}
V^{oge~\Lambda_{Q^*}}_{so(dynamic)} = {1 \over m_Q \mu_{\sigma Q}(\sqrt{3 \over 2}
\lambda)}
{dV^{eff}_{Coulomb} \over d(\sqrt{3 \over 2} \lambda)} \vec L_{\lambda}
\cdot \vec S_Q
\end{equation}
as in Eq. (\ref{eq:VsoLambdaQ}).  This completes our check that in the
$\Lambda_{Q^*}$
states one-gluon-exchange
spin-orbit forces are purely meson-like.

	In the absence of a microscopic picture of confinement, I do not
know how to improve on the
argument made earlier that since confinement arises out of the color
electric field,
Thomas precession in the confining potential will be that induced by a
confining force directed
along $\vec \lambda$, in the same direction as $\vec E_Q$ in Eq.
(\ref{eq:EQ}).
It should be noted that any microscopically
two-body model for confinement of the form
\begin{equation}
V_{conf} = \sum_{i<j} V_{conf}^{ij}(r_{ij})
\end{equation}
will lead to this result since then the force on $Q$ from particle $1$ will be
\begin{equation}
\vec F_{1Q}= -{1 \over r_{1Q}}
{dV^{1Q}_{conf} \over dr_{1Q}} \vec r_{1Q}~~.
\end{equation}
With, $\vec r_{1Q} = \sqrt \frac{3}{2} \vec \lambda + \sqrt \frac {1}{2}
\vec \rho$, in a
$\Lambda_{Q^*}$ state only the $\sqrt \frac {3}{2} \vec \lambda$
term survives since the $\vec \rho$ integration can be done over annular
regions
of fixed $\vert \vec r_{1Q} \vert$.  (The Coulomb potential is thus only
special among two-body forces in that
its dependence on $\lambda$
is dictated by Gauss' Law, not that it leads to a force directed along
$\vec \lambda$.)  Given its
implications for the relationship between the conclusions drawn here about
spin-orbit forces and those
in the literature, it is also useful to explicitly consider the case of ``harmonic
confinement" in which
$V^{ij}_{conf}={1 \over 2}kr^2_{ij}$.
In this case the total confining force on $Q$ is always $\sqrt{6} k \vec
\lambda$
{\it independent of the state of the $\vec \rho$
variable}, so the spin-orbit forces on $Q$ from Thomas precession in a
harmonic potential
are {\it always}
purely quasi-two-body.

	To summarize, we have found and confirmed that, as expected,
the $\Lambda_{Q^*}$ baryons have no spin-orbit puzzle,
and in particular that they have only meson-like quasi-two-body spin-orbit
forces.  This conclusion
implies that the three-body spin-orbit forces described in the literature
conspire in the case of
$\Lambda_{Q^*}$ baryons to produce quasi-two-body behaviour.
(One may readily check using the matrix elements provided in
Ref. \cite{IK} that this is true for $\ell_\lambda=1$.)
We will discuss this new perspective on the
baryon spin-orbit puzzle below.

\subsection {The $\Lambda_{Q^*}$ Baryons: Phenomenology}

    The preceding discussion of spin-orbit forces in $\Lambda_{Q^*}$
baryons is important
conceptually in defining a tower of baryon excitations with no
spin-orbit problem and in creating a new perspective on the spin-orbit puzzle.
However,
it does not have much useful predictive power. In the $\Lambda_{s^*}$'s, where
there is substantial data, the two-body spin-orbit force is (as in the
$K^*$'s) the ill-determined difference of a large positive dynamical spin-orbit term
and a large negative Thomas precession term, while in the heavy quark sectors
$\Lambda_{c^*}$ and $\Lambda_{b^*}$ the simple dynamical term dominates so
that the predictions are reasonably reliable, but there is little data
available.  The calculations themselves are straightforward, requiring only
the
matrix elements of $V_{so} ^{\Lambda_{Q^*}}$ of Eq. (\ref{eq:VsoLambdaQ})
in the wave functions
$\psi_{0\ell_\lambda\ell_\lambda}$ of Eq. (\ref{eq:psilambda}).  
It should be noted that the required matrix elements
(especially those of $1/r^3$) are not very accurately determined 
by these harmonic oscillator wave functions. This accuracy (which is typically about $25\%$
in low-lying states)
could easily be improved with better wave functions, 
the most significant qualitative aspect of such an improvement program 
being an increase of the $1/r^3$ relative to the $1/r$ matrix elements.
However, the value of such an improvement is dubious:
based on the expectation value of $E/m$, the nonrelativistic framework
of this entire discussion can only be expected to be accurate at the $\pm 50\%$ level.
I therefore present in Table II the separated dynamical and Thomas precession
spin-orbit contributions to emphasize not the numerical results, but
rather the cancellation that is at the heart of the solution of
the spin-orbit puzzle. (Note that while the matrix elements of $1/r^3$ and
$1/r$ are decreasing with $\ell$, since $\langle \vec L_\lambda \cdot \vec S_Q \rangle $ grows,
both contributions remain substantial through $\ell=4$.) By assigning theoretical errors of $\pm 50\%$
to each contribution, the {\it quantitative} reliability of
the various predicted splittings can also be assessed.

\bigskip\bigskip\bigskip

\begin{table}
  \caption { The predicted spin-orbit splittings in $\Lambda_{Q^*}$
baryons for $\ell=$1, 2, 3, and 4 (in MeV), shown in the format
$\Delta E_{dynamical}+\Delta E_{Thomas}$. Theoretical errors may be
estimated by allowing each term to vary by $\pm 50\%$. Experimental
splittings are
given in parentheses below the predictions when they are known.}
\vspace{0.4cm}
\begin{center}
  \begin{tabular}{|cccccc|}
&$\ell$  & $\Delta m_{\Lambda}$ & $\Delta m_{\Lambda_c}$ & $\Delta
m_{\Lambda_b}$ & \\
\hline
&$1 \leftrightarrow {3 \over 2}^- - {1 \over 2}^-$ & $221-175$  & $75-23$  &
$26-3$   & \\
&& ($112 \pm 5$) & ($33 \pm 1$) & & \\
&$2 \leftrightarrow {5 \over 2}^+ - {3 \over 2}^+$ & $118-246$  & $39-21$  &
$13-3$   & \\
&& ($-60 \pm 30$) && & \\
&$3 \leftrightarrow {7 \over 2}^- - {5 \over 2}^-$ & $81-266$  & $26-21$  &
$9-3$   & \\
&& ($-10 \pm 30$) && & \\
&$4 \leftrightarrow {9 \over 2}^+ - {7 \over 2}^+$ & $42-199$  & $19-23$  &
$7-3$   & \\
&&&& & \\
  \end{tabular}
 \end{center}
\end{table}

\bigskip\bigskip

\section {Conclusions}
	Figure 1(b) suggests that the tower of $\Lambda_{Q^*}$ baryons I have
described here will provide the simplest setting in which to begin to
understand
the baryon spin-orbit problem since, at least naively, they ought to have
meson-like dynamics.  I have shown here that this expectation is in fact
correct and that this select tower of baryons has no ``spin-orbit puzzle".
The key to the solution of the puzzle for this select group of baryons is
that for
them the microscopic three-body spin-orbit forces described in the
literature conspire to produce only meson-like quasi-two-body spin-orbit
forces.
As a result these states experience the same strong cancellation between
dynamical spin-orbit forces and Thomas precession which leads to small
spin-orbit effects in mesons.

	It is possible that this work has no implications beyond
identifying these
special states.  However, in studying
these states we have also developed a new perspective on the forces
in baryons which may have a wider utility in confronting the baryon
spin-orbit puzzle.  In particular, we have seen that the standard
classification
in the literature of spin-orbit forces as being two- and three-body is somewhat
misleading.  This classification is correct at the microscopic level
(see Eq. (\ref{eq:3bodyso})), but not very relevant to the issue of whether
the spin-orbit forces on a given quark $Q$ appear to arise from the baryon
center-of-mass, i.e., whether they are quasi-two-body in character.  With the
spin-orbit puzzle solved for a slice through the baryon spectrum, it would
be odd
if there were an intractable spin-orbit problem in baryons. At the least,
one may
hope that the
simple interpretation  described here in terms of quasi-two-body (i.e.,
meson-like)
dynamics will lead to further insights into the spin-orbit puzzle.

    In fact the arguments presented here for the $\Lambda_{Q^*}$ baryons
are immediately generalizable to the spin-orbit force on the third quark $Q$
in  $^2\Sigma_{Q^*}$ and $^4\Sigma_{Q^*}$ states, the total
quark spin $1/2$ and $3/2$ $\Sigma_Q$ baryons with the same spatial
wave functions as $\Lambda_{Q^*} \equiv ~ ^2\Lambda_{Q^*}$.
Since these three ``uds basis" states \cite{Franklin,IK} are sufficient to
completely describe the SU(3) baryon spectrum, it might seem that an 
even more substantial generalization is in hand: in the SU(3) limit, baryon 
wave functions have permutation symmetry, so proving that quark 3 has meson-like spin-orbit
forces is sufficient to prove that each quark does.
However, in the SU(3) limit $\omega_{\rho}=\omega_{\lambda}$ and our 
proofs are inapplicable: with such a
degeneracy, leading order spin-orbit effects can occur {\it via} mixing
between $\vec \rho$ and $\vec \lambda$ excitations. Thus extending 
the considerations presented here beyond the spin-orbit force
on $Q$ in   $^2\Lambda_{Q^*}$, $^2\Sigma_{Q^*}$, and $^4\Sigma_{Q^*}$ 
will be nontrivial. Using our new perspective of quasi-two-body forces, we can visualize at
least one obstacle to such extensions. Consider the
counterparts to
our $\Lambda_{Q^*}$ state in which the $\vec \rho$ variable is in a state with
$\ell_{\rho}\not=0$ but $\ell_{\lambda}=0$. In these circumstances
the $ud$ pair will
develop an orbital color magnetic moment proportional to $\vec L_\rho$ with
which the color magnetic moment $\vec S_Q/m_Q$ will in general interact.

   It would thus seem to require a careful
reanalysis of
the effects of spin-orbit splittings on the baryon spectrum from this new
perspective to determine if all of the {\it observed} nearly degenerate
spin-orbit multiplets can be accommodated within it, and, if they
cannot be, to define the next steps along the path to solving this old problem.
It of course remains possible, despite the tantalizing fact that mesons and now
the $\Lambda_{Q^*}$ states admit a nonrelativistic solution to their spin-orbit
splittings, that the ultimate solution is as suggested in
Ref.\cite{CapstickI}: that
relativistic effects have simply produced a gross enhancement of spin-spin over spin-orbit interactions.

\vfill\eject

{\centerline {\bf ACKNOWLEDGEMENTS}}

\medskip

    This work was supported by DOE contract DE-AC05-84ER40150 under which the
Southeastern Universities Research Association (SURA) operates
the Thomas Jefferson National Accelerator Facility.

\bigskip\bigskip

{

\end{document}
\begin{references}
\bibitem{oporder} There is no operator ordering ambiguity between $\vec p_Q$
and $\vec r_{iQ}/r^3_{iQ}$ because they appear in a cross-product.
\end{references}
